\documentclass[english,10pt]{article}
\usepackage[papersize={19cm,25cm},body={15cm,21cm},centering]{geometry}

\usepackage[dvipsnames]{xcolor}
\usepackage{tikz}
\usetikzlibrary{shapes,calc,matrix}
\usepackage[british]{babel}
\usepackage[utf8]{inputenc}
\usepackage[normalem]{ulem}
\usepackage[T1]{fontenc}
\usepackage{amsmath}
\usepackage{amsthm}
\usepackage{mathtools}
\mathtoolsset{showonlyrefs=true}

\usepackage{url}
\usepackage{amsfonts}
\usepackage{amssymb}
\usepackage{graphicx}
\usepackage{caption} 
\usepackage{subcaption}
\usepackage{enumitem}
\usepackage{bm}
\usepackage{stfloats}
\usepackage[final]{pdfpages}
\usepackage{slashbox}
\usepackage{ulem}
\usepackage{systeme}
\usepackage{algorithm, float}
\usepackage{algorithmic} 
\usepackage[squaren, Gray]{SIunits}
\usepackage[acronym,toc,nonumberlist, nopostdot, style=super, nogroupskip]{glossaries}
\usepackage{hyperref}
\usepackage[figurename=Fig.]{caption}
\newtheorem{lemma}{Lemma}
\newcommand{\algorithmicbreak}{\textbf{break}}
\newcommand{\BREAK}{\STATE \algorithmicbreak}

\DeclareMathOperator*{\argmin}{arg\,min}

\renewcommand{\algorithmicrequire}{\textbf{Input:}}

\begin{document}
\title{Near-Convex Archetypal Analysis}
\date{}
\author{
Pierre De Handschutter \quad 
Nicolas Gillis \quad Arnaud Vandaele \quad Xavier Siebert\thanks{This work was supported by the European Research Council (ERC starting grant n$^\text{o}$ 679515), and 
by the Fonds de la Recherche Scientifique - FNRS and the Fonds Wetenschappelijk Onderzoek - Vlanderen (FWO) under EOS Project no O005318F-RG47. 
E-mails: \{pierre.dehandschutter,
nicolas.gillis,
arnaud.vandaele,
xavier.siebert\}@umons.ac.be.} \\
Department of Mathematics and Operational Research \\
Facult\'e Polytechnique, Universit\'e de Mons \\
Rue de Houdain 9, 7000 Mons, Belgium
}

\maketitle
\begin{abstract} 
Nonnegative matrix factorization (NMF) is a widely used linear dimensionality reduction technique for nonnegative data. 
NMF requires that each data point is approximated by a convex combination of basis elements. 
Archetypal analysis (AA), also referred to as convex NMF, 
is a well-known NMF variant  
imposing that the basis elements are themselves convex combinations of the data points. 
AA has the advantage to be more interpretable than NMF because the basis elements are directly constructed from the data points. 
However, it usually suffers from a high data fitting error because the basis elements are constrained to be contained in the convex cone of the data points. 
In this letter, we introduce near-convex archetypal analysis (NCAA) which combines the advantages of both AA and NMF. 
As for AA, the basis vectors are required to be linear combinations of the data points and hence are easily interpretable. 
As for NMF, the additional flexibility in choosing the basis elements allows NCAA to have a low data fitting error.   
We show that NCAA compares favorably with a state-of-the-art minimum-volume NMF method on synthetic datasets and on a real-world hyperspectral image. 
\end{abstract}


\section{Introduction}
%
%

Nonnegative Matrix Factorization (NMF) is a well-known technique in unsupervised data analysis; see for example~\cite{gillis2014and, xiao2019uniq} and the references therein. 
Given an $m$-by-$n$ nonnegative input matrix 
$X \in \mathbb{R}^{m  \times n}_+$ and a factorization rank $r$, 
the goal of NMF is to find two nonnegative matrices $W \in \mathbb{R}^{m  \times r}_+$ and 
$H \in \mathbb{R}^{r  \times n}_+$ such that $X \approx WH$.  
The standard NMF optimization problem is formulated as follows 
\begin{equation}
\label{NMF}
\underset{\substack{W \in \mathbb{R}^{m  \times r}, H \in \mathbb{R}^{r  \times n}}}{\min} \Vert X-WH \Vert_F^2 \text{ such that } W \geq 0 \text{ and } H \geq 0, 
\end{equation} 
where $||A||_F^2 = \sum_{i,j} A(i,j)^2$ is the squared Frobenius norm of matrix $A$. 
The matrix $W$ is referred to as the matrix of basis elements while the matrix $H$ indicates the proportions in which each basis vector is present in any data point. 
Due to its physical interpretation, for example in hyperspectral unmixing (see Section~\ref{sec:blindhu}), the matrix $H$ is often required to have the sum of the entries of each column less or equal to $1$. 
With this constraint and denoting $\Delta^r =\big\lbrace x \in \mathbb{R}^r | x \geq 0, \sum\limits_{i=1}^r x_i \leq 1 \big\rbrace$, we consider in this paper the following problem 
\begin{equation}
\label{NMFv2}
\underset{\substack{W \in \mathbb{R}^{m  \times r}_+ \\ H \in \mathbb{R}^{r  \times n}_+}}{\min} \Vert X-WH \Vert_F^2 \text{ such that } H(:,j) \in \Delta^r \text{ for all } j.  
\end{equation} 
A notable variant of NMF is archetypal analysis (AA)~\cite{cutler1994archetypal}. 
In AA, an additional constraint imposes that the basis vectors, referred to as archetypes, are themselves convex combinations of the data points, that is $W(:,k)=XA(:,k)$ with $A(:,k)\in \Delta^n$, $k=1,...,r$. The problem becomes 
\begin{equation}
\label{thisisit}
\begin{gathered}
\underset{\substack{
A(:,k) \in \Delta^n \text{ for }  k=1,\dots,r \\ 
H(:,j) \in \Delta^r \text{ for }  j=1,\dots,n
}
}{\min} \Vert X-XAH \Vert_F^2. 
\end{gathered}
\end{equation} 
AA has also been introduced under the name convex NMF~\cite{ding2008convex}. 
While AA offers higher guarantees of interpretability since the archetypes have to belong to the convex hull of the data points,  
the reconstruction error is likely to be higher than in NMF. 
Hence the membership of the archetypes to the convex hull of the columns of $X$ was relaxed in some previous works. 
In~\cite{morup2012archetypal}, the sum of the entries of each column of $A$ is allowed to be between $1-\delta$ and $1+\delta$ for some $\delta \geq 0$ fixed \textit{a priori}. 
In~\cite{javadi2019non}, the authors combine AA and NMF through a trade-off between the reconstruction error and the distance between the archetypes and the convex hull of $X$. However, the variables involved are the NMF ones, namely $W$ and $H$. 
Minimum-volume NMF~\cite{miao2007endmember} is an NMF variant that minimizes the volume of the convex hull of the columns of $W$; see~\cite{xiao2019uniq} and the references therein for details. 
The latter two approaches, though close in spirit to AA, do not allow to interpret how the archetypes are built from the data through a coefficient matrix $A$.  

In this work, we propose a new model, dubbed near-convex archetypal analysis (NCAA), which benefits from the advantages of both NMF via low reconstruction error and AA via interpretability. 
This letter is organized as follows. In Section~\ref{models}, we state the model and explain its geometric interpretation. 
We detail the optimization framework in Section~\ref{opti}. 
We present the performances of our algorithm on both synthetic and real datasets in Section~\ref{exp} and draw some possible perspectives of future research in Section~\ref{conclusion}.

\section{Near-convex archetypal analysis} \label{models}

Given a data matrix $X \in \mathbb{R}^{m  \times n}$, 
a matrix $Y \in \mathbb{R}^{m  \times d}$ with $d$ columns,  
the rank $r$ of the factorization 
and a positive scalar $\epsilon$, 
we define the NCAA problem as follows 
\begin{equation} \label{my_model}
\begin{gathered}
\min_{
\substack{A \in \mathbb{R}^{d\times r} 
\\ 
H(:,j) \in \Delta^r \text{ for }  j=1,\dots,n 
}
} 
\Vert X- YAH\Vert^{2}_{F} \\
\hspace{-17mm} \text{such that} \quad A(k,l) \geq - \epsilon \quad \text{for all} \; k,l, \quad \epsilon \geqslant 0, \\
\sum\limits_{k=1}^d A(k, l) = 1 \quad \text{for all} \; l =1, \dots, r, 
\end{gathered}
\end{equation}
For $Y = X$ and $\epsilon = 0$, NCAA~\eqref{my_model} coincides with AA~\eqref{thisisit}. 
Let us point out the two differences between NCAA and AA:   
\begin{itemize} 

\item  For $\epsilon > 0$, the archetypes are allowed to lie outside the convex hull of the data points and are said to be near-convex combinations (NCCs) of the data points.

\item In order to reduce the computational cost compared to AA, the basis vectors $YA$ are combinations of a matrix $Y$, made of $d$ points such that $d \ll n$. In practice (see below), we will choose these $d$ columns as a subset of the columns of $X$. Note that, in~\cite{bauckhage2014note}, $Y$ is chosen as the vertices of the convex hull of $X$. However, in noisy scenarios, most data points will be vertices hence this approach usually does not allow to have $d$ significantly smaller than $n$.  

\end{itemize}

NCCs have an interesting geometric interpretation, as stated in the following lemma.  
\begin{lemma} 
\label{my_lemma}
Let $Y \in \mathbb{R}^{m \times d}$, and let us define the columns of the matrix $Z$ as 
\begin{equation}
Z(:,j) = Y(:,j) + d \epsilon (Y(:,j)-\overline{y}) \text{ for } j=1,\dots,d, 
\end{equation}
where $\overline{y}$ is the average of the columns of $Y$, that is, $\overline{y} = Ye/d$ where $e$ is the vector of all ones.
Then, the NCC of the columns of $Y$, that is, the set 
$\mathcal{Y} = 
\big\{ x 
\ | \ x = Ya, 
\sum_i a_i = 1, a_i \geq -\epsilon \big\}$, 
is equal to the convex combinations of the columns of $Z$, that is, to the set 
$\mathcal{Z} =
\big\{ x 
\ | \ x = Za, 
\sum_i a_i = 1, a_i \geq 0  \big\}$. 
\end{lemma}
Fig.~\ref{Geom_interp} illustrates the geometric interpretation given in Lemma~\ref{my_lemma} for $d=2r=6$: each $Z(:,j)$ is aligned with the corresponding $Y(:,j)$ and $\overline{y}$, and lies outside the convex hull of the columns of $Y$. 

\begin{center} 
\begin{figure}
\setlength{\belowcaptionskip}{0pt}
\begin{center}
\begin{tikzpicture}[scale=1,yscale=1]
	\tikzstyle{dot}=[diamond,minimum size=5pt,inner sep=0pt,outer sep=-1pt]
	\tikzstyle{carre}=[minimum size=4pt,inner sep=0pt,outer sep=-1pt]
	\tikzstyle{cercle}=[circle,minimum size=4pt,inner sep=0pt,outer sep=-1pt]
    	\tikzstyle{important line}=[very thick]
    	\tikzstyle{information text}=[rounded corners,fill=red!10,inner sep=1ex]

    	\foreach \p in {1,...,1000}
    	{ \pgfmathsetmacro{\x}{2.7*rand}
    	    \pgfmathsetmacro{\y}{2.2*rand}
    	    	\xdef\ok{1}
    	    	\pgfmathparse{3*\y+4*\x < -9 ? int(1) : int(0)}
    	    	\ifnum\pgfmathresult=1 
    	    	\xdef\ok{0}
    	    	\fi
    	    	\pgfmathparse{3*\y-4*\x < -9 ? int(1) : int(0)}
    	    	\ifnum\pgfmathresult=1 
    	    	\xdef\ok{0}
    	    	\fi
    	    	\pgfmathparse{\y+\x > 3.7 ? int(1) : int(0)}
    	    	\ifnum\pgfmathresult=1 
    	    	\xdef\ok{0}
    	    	\fi
    	    	\pgfmathparse{\y-\x > 3.7 ? int(1) : int(0)}
    	    	\ifnum\pgfmathresult=1 
    	    	\xdef\ok{0}
    	    	\fi
    	    	\pgfmathparse{\y > 1.5 ? int(1) : int(0)}
    	    	\ifnum\pgfmathresult=1 
    	    	\xdef\ok{0}
    	    	\fi
				\ifnum\ok=1
				\fill[blue]    (\x,\y) circle (0.025);
				\fi  
   		 }
   		 
   	\node[star,fill=black,minimum size=5pt,inner sep=0pt,outer sep=-1pt] at (0,0) (ybar) {};
    	
    	\node[carre,fill=Green!80] at (4,1)   (Z4) {}; \node[Green!80] at (3.6,2.4) (Z4t) {\scriptsize $Z(:,5)$};
    	\node[carre,fill=Green!80] at (3,2)   (Z5) {}; \node[Green!80] at (4.7,1.1) (Z5t) {\scriptsize $Z(:,4)$};
    	\node[carre,fill=Green!80] at (-4,1)  (Z1) {}; \node[Green!80] at (-4.7,1.1) (Z1t) {\scriptsize $Z(:,1)$};
    	\node[carre,fill=Green!80] at (-3,2)  (Z6) {}; \node[Green!80] at (-3.6,2.4) (Z6t) {\scriptsize $Z(:,6)$};
    	\node[carre,fill=Green!80] at (-1,-3) (Z2) {}; \node[Green!80] at (-1.4,-3.4) (Z2t) {\scriptsize $Z(:,2)$};
    	\node[carre,fill=Green!80] at (1,-3)  (Z3) {}; \node[Green!80] at (1.4,-3.4) (Z3t) {\scriptsize $Z(:,3)$};
    	
    	\node[cercle,fill=red] (W1) at ($(Z1)!0.5!(Z6)$) {}; \node[red] at (-4.2,1.7) (W1t) {\scriptsize $W(:,1)$};
    	\node[cercle,fill=red] (W2) at ($(Z4)!0.5!(Z5)$) {}; \node[red] at (4.2,1.7) (W2t) {\scriptsize $W(:,2)$};
    	\node[cercle,fill=red] (W3) at ($(Z2)!0.5!(Z3)$) {}; \node[red] at (0,-3.4) (W3t) {\scriptsize $W(:,3)$};

	\def\f1{1-(24/29)}
    	\node[dot,fill=Maroon!80] (Y5) at ($(W2)!\f1!(W1)$) {}; \node[Maroon!80] at (1.7,1.7) (Y5t) {\scriptsize $Y(:,5)$};
    	\node[dot,fill=Maroon!80] (Y6) at ($(W1)!\f1!(W2)$) {}; \node[Maroon!80] at (-1.7,1.7) (Y6t) {\scriptsize $Y(:,6)$};
    	\def\f2{1-(24/29)}
    	\node[dot,fill=Maroon!80] (Y4) at ($(W2)!\f2!(W3)$) {}; \node[Maroon!80] at (3.2,0.3) (Y4t) {\scriptsize $Y(:,4)$};
    	\node[dot,fill=Maroon!80] (Y3) at ($(W3)!\f2!(W2)$) {}; \node[Maroon!80] at (1.3,-2.3) (Y3t) {\scriptsize $Y(:,3)$};
    	\def\f3{1-(24/29)}
    	\node[dot,fill=Maroon!80] (Y2) at ($(W3)!\f3!(W1)$) {}; \node[Maroon!80] at (-1.25,-2.3) (Y2t) {\scriptsize $Y(:,2)$};
    	\node[dot,fill=Maroon!80] (Y1) at ($(W1)!\f3!(W3)$) {}; \node[Maroon!80] at (-3.2,0.3) (Y6t) {\scriptsize $Y(:,1)$};

    	\draw[cyan,dashed,very thick] (Z2) -- (Z3) -- (Z4) -- (Z5) -- (Z6) -- (Z1) -- (Z2);
    	\draw[orange,very thick] (Y2) -- (Y3) -- (Y4) -- (Y5) -- (Y6) -- (Y1) -- (Y2);
    	
    	\draw[black,dashed,very thick] (Z1) -- (ybar); \node[black] at (0.2,0.2) (Z4t) {\small $\bar{y}$};
    	\draw[black,dotted,very thick] (W1) -- (W2);
    	\draw[black,dotted,very thick] (W1) -- (W3);
    	\draw[black,dotted,very thick] (W2) -- (W3);
    	
    	\matrix[draw, row sep=-0.1cm, column sep=0.3pt] at (5.3,-2) {
  	\node[circle,fill=blue,label=right:{Data points $X$},minimum size=2pt,inner sep=0pt,outer sep=0pt] {}; \\
  	\node[dot,fill=Maroon!80,label=right:{Points $Y$}] {}; \\
  	\node[carre,fill=Green!80,label=right:{Points $Z$}] {}; \\
  	\node[cercle,fill=red,label=right:{Basis vectors $W$}] {}; \\
  	\node[star,fill=black,label=right:{Mean vector $\bar{y}$},minimum size=5pt,inner sep=0pt,outer sep=0pt] {}; \\
	};
    		\end{tikzpicture}
    		\end{center}
\caption{Geometric interpretation of NCAA for $r=3$, $d=2r=6$, $m=2$: as $\epsilon$ grows, the estimated basis vectors are further away from the convex hull of $X$. In this example, $p=0.8$ and $\epsilon=\frac{1}{14}$. \label{Geom_interp}}
\end{figure}
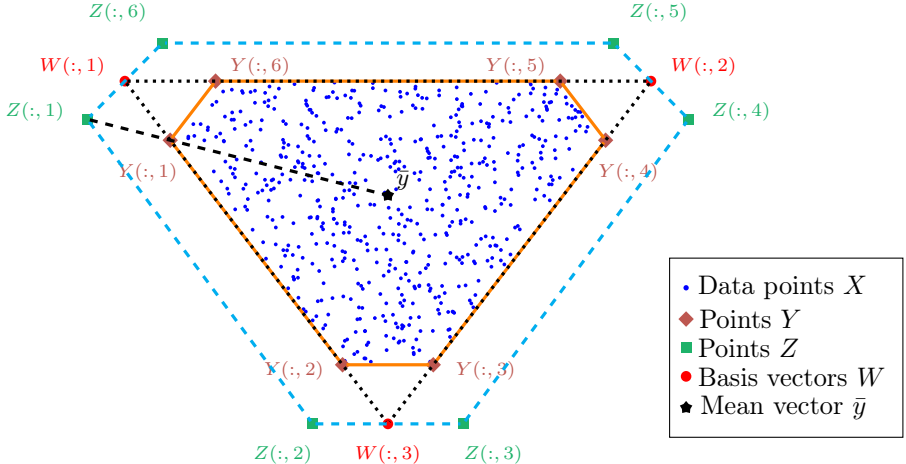
\end{center}

This interpretation is rather interesting: as $\epsilon$ increases, the archetypes $W = YA$ are allowed to lie further away from the convex hull of the columns of $Y$. Let us define the purity level $p$ of $X = WH$ as $p 
= \min_{1 \leq i \leq r} \max_{1 \leq j \leq n} H(i,j)$. In Fig.~\ref{Geom_interp}, $p = 0.8$. 
If $p=1$, the data is said to be separable (see for example~\cite{xiao2019uniq}), as the basis vectors $W$ correspond to some of the points in $X$ so that $\epsilon$ can be chosen equal to 0. 

There are two key aspects in the NCAA model: the choice of $Y$ and the choice of $\epsilon$. 
The value of $\epsilon$ will be tuned automatically within the algorithm; see Section~\ref{opti} for more details.  
For the choice of $Y$, we use two simple schemes but others could be considered:  

\begin{itemize} 
\item Successive nonnegative projection algorithm (SNPA)~\cite{gillis2014successive}, designed to solve separable NMF, extracts extreme points of the dataset but is sensitive to outliers (although appropriate pre- and post-processing could resolve this issue). 

\item Hierarchical clustering (HC)~\cite{gillis2015hierarchical}, designed to cluster data points in a hierarchical way, identifies points that are not necessarily extreme points of the data cloud but is less sensitive to outliers hence appropriate for real data sets.
\end{itemize}

The number of points $d$ in $Y$ is chosen \textit{a priori} depending on the application (so is the rank $r$ as in most NMF models). It can be chosen as a small multiple of $r$:  typically, a value between $2r$ and $10r$ works well in practice; see the numerical experiments for some examples.

\section{Optimization framework} \label{opti}

We propose a standard optimization framework to solve~\eqref{my_model}, namely a two-block coordinate descent (BCD) \cite{kim2014algorithms}. 
It consists in alternatively optimizing $A$ and $H$ keeping the other fixed; this is the standard framework for most NMF algorithms~\cite{gillis2014and}. 
The optimization of $A$ and $H$ is performed through a fast projected gradient descent method (FPGM) with Nesterov acceleration~\cite{nesterov1983method}. 
The step size is tuned with a backtracking line search. 
The matrix $H$ is projected onto the unit simplex through the algorithm described in Appendix of~\cite{gillis2014successive}. 
The projection of $A$ can be performed in a similar way but requires the implementation of an efficient column-wise algorithm, valid for any $\epsilon$. 
 We have implemented such an approach; see the Matlab code available from \begin{center}
 \url{http://bit.ly/NCAAv1} 
 \end{center}
   \begin{algorithm} [h]
        \caption{NCAA} \label{algo1}
 \begin{algorithmic}[1]
 \renewcommand{\algorithmicrequire}{\textbf{Input:}}
  \REQUIRE Nonnegative matrices 
  $X\in \mathbb{R}_{+}^{m \times n}$ and 
  $Y\in \mathbb{R}_{+}^{m \times d}$, 
  rank $r$, 
  bounds 
  $0 < \epsilon_{\min} < \epsilon_{\max}$, 
  tolerance $\delta$ \\
 \renewcommand{\algorithmicrequire}{\textbf{Output:}}
  \REQUIRE Matrices 
  $A\in \mathbb{R}^{d \times r}$ and 
  $H\in \mathbb{R}^{r \times n}$ 
  that solve~\eqref{my_model} \\
 
 \STATE Compute initial matrices $A^{(0)}$ and $H^{(0)}$, $i=0$, $\epsilon^{(1)}=\epsilon_{\min}$. 

\STATE   err$(0)= ||X- Y A^{(0)} H^{(0)}||_F^2$
  
  \FOR {$t = 1, 2,\dots$}
  \FOR {$k = 1, 2,\dots,50$}
  \STATE $i=i+1$
  \STATE $A^{(i)} = \underset{
\substack{A(:,l) \in \Delta_{\epsilon^{(t)}}^d \forall \; l}
} \argmin
\Vert X- YAH^{(i-1)}\Vert^{2}_{F} $ ($\star$)
\\
  \STATE $H^{(i)} = \underset{
\substack{H(:,j) \in \Delta^r \forall \; j}
} \argmin
\Vert X- YA^{(i)}H\Vert^{2}_{F}$  ($\star$)\\
  \STATE err$(i)= ||X- Y A^{(i)} H^{(i)}||_F^2$ \\
  \ENDFOR
  
   \IF {err$(i)-$err$(i-1) < \delta$ err$(0)$}
   \STATE $\epsilon_{\max}=\epsilon^{(t)}$; 
   $\epsilon^{(t+1)}=\frac{\epsilon_{\min}+\epsilon_{\max}}{2}$
   \ELSE 
    \STATE $\epsilon_{\min}=\epsilon^{(t)}$; 
    $\epsilon^{(t+1)}=\min(2\epsilon^{(t)}, \epsilon_{\max})$
   \ENDIF
  
    \ENDFOR
    
  $\star$ The problem is solved via FPGM. 
 \end{algorithmic}
 \end{algorithm}

The value of the parameter $\epsilon$ is tuned automatically as it highly depends on the data distribution. We believe this is a strong advantage of our method. 
For example, in the separable case, that is, when the basis vectors belong to the data points~\cite{arora2016computing}, $\epsilon$ should be set to $0$. 
However, when this is not the case, 
$\epsilon$ should be chosen more carefully; see for example Fig. \ref{Geom_interp} for a non-separable case. 
We proceed as follows. The value of $\epsilon$ is initially set to a very small value (we have used $\epsilon_{\min}=10^{-3}$ in all numerical experiments) which imposes that the archetypes are close to the convex hull of $X$. Then, $\epsilon$ is doubled at each iteration as long as the relative error decreases by a given tolerance between two consecutive iterations (in our implementation we have used $\delta = 10^{-4}$). 
Intuitively, the idea is to slowly allow the basis vectors to lie further away from the convex hull of $X$. 
This tuning process is described in Algorithm~\ref{algo1}. We will denote $\Delta_{\epsilon}^r =\big\lbrace x \in \mathbb{R}^r | x \geq -\epsilon \bm{1}, \sum\limits_{i=1}^r x_i \leq 1 \big\rbrace$.

In the model~\eqref{my_model} and in Algorithm~\ref{algo1}, the value of $\epsilon$ is the same for all entries of $A$. However, in practice, it may be crucial to allow the columns of $W$ to be closer or further away from the data points. For example, it may happen that a subset of the archetypes belong to the data points (that is, some columns of $W$ appear as columns of $Y$) for which the value of $\epsilon$ should be equal to zero. 
Therefore, rather than considering a unique $\epsilon$ in the model, we impose instead that 
\[
A(:,l) \geq -\epsilon_l \quad \text{ for } \quad  l = 1,\dots,r, 
\]
where $\epsilon_l \geq 0$ ($1 \leq l \leq r$) are parameters. 
To deal with this non-symmetric case, we propose a fine-tuning stage, after Algorithm~\ref{algo1} has terminated. 
Starting from the value of $\epsilon$ computed by Algorithm~\ref{algo1}, 
$\epsilon_l$ is fine-tuned for each basis vector $YA(:,l)$, one at a time and independently of each other (keeping the others fixed at the value returned by Algorithm~\ref{algo1}). It works as follows. The value of $\epsilon_l$ ($l=1,\dots,r$) is decreased by a factor $\alpha$ (set to $0.8$ in the implementation) until the reconstruction error becomes $1\%$ larger than the error at the end of the global tuning; see Algorithm~\ref{algo2}. Intuitively, we move back each column of $W = YA$ towards the convex hull of the columns of $Y$ as long as the error does not increase too much.  

\begin{algorithm}
 \caption{Fine tuning of NCAA} \label{algo2}
 \begin{algorithmic} [t]
 \renewcommand{\algorithmicrequire}{\textbf{Input:}}
  \REQUIRE Nonnegative matrices 
  $X\in \mathbb{R}_{+}^{m \times n}$, 
  $ Y\in \mathbb{R}_{+}^{m \times d}$,  
  $A\in \mathbb{R}^{d \times r}$, 
  $H\in \mathbb{R}^{r \times n}$, rank $r$, tolerance $\delta^*$\\
 \renewcommand{\algorithmicrequire}{\textbf{Output:}}
  \REQUIRE Matrices $A^{(*)} \in \mathbb{R}^{d \times r}$, $H^{(*)} \in \mathbb{R}^{r \times n}$ that solve \eqref{my_model} \\
   \STATE $\text{err}_0 = ||X-YAH||_F^2$ \\
  \FOR {$l = 1,\dots,r$}
  \STATE $B=A$ \\
  \STATE $\epsilon_l^{(0)}=-\min(A(:,l))$ \\
   \FOR {$t = 1, 2,\dots$}
  \STATE $\epsilon_l^{(t)}=\alpha \epsilon_l^{(t-1)}$ \\
  \FOR {$k = 1, 2,\dots,50$}
  \STATE $i=i+1$
  \STATE $B^{(i)}(:,l) = \underset{
\substack{B(:,l) \in \Delta_{\epsilon_l^{(t)}}^d}
} \argmin
\Vert X- YBH^{(i-1)}\Vert^{2}_{F} $ ($\star$)\\
  \STATE $H^{(i)} = \underset{
\substack{H(:,j) \in \Delta^r \forall \; j}
} \argmin
\Vert X- YB^{(i)}H\Vert^{2}_{F}$  ($\star$)\\
  \STATE err$(i)=||X- Y B^{(i)} H^{(i)}||_F^2 $ \\
  \ENDFOR

   \IF {$\text{err}(i)>\delta^* \text{err}_0$}
   \STATE $A^{(*)}(:,l)=B^{(i)}(:,l)$ \\
   \BREAK
   \ENDIF
  
    \ENDFOR
     \ENDFOR
     \STATE $H^{(*)} = \underset{
\substack{H(:,j) \in \Delta^r \forall \; j}
} \argmin
\Vert X- YA^{(*)}H\Vert^{2}_{F}$  ($\star$)\\ 

 $\star$ The problem is solved via FPGM. 
 \end{algorithmic}
 \end{algorithm}

\paragraph{Computational cost}   
The main computational cost of Algorithms~\ref{algo1} and~\ref{algo2} lies in the computation of the gradient and the projection, which both require matrix-matrix multiplications. One can check that the computational cost per iteration is $\mathcal{O}(mnr)$ operations.  As long as $d$ is chosen small enough (we recommend a multiplicative factor of $r$), 
the computational cost of the algorithm remains linear in the dimensions of the input matrix as for classical NMF, hence can be applied to large-scale data sets. 
Note that the number of variables of NCAA is $r(d+n)$ which is less than the $2nr$ of the classical AA but of the same order as the number of variables in NMF $r(m+n)$, as $d$ is usually smaller than $m$.

\section{Numerical experiments} \label{exp}

In this section, the performances of NCAA are evaluated on synthetic data sets, and on a real-world hyperspectral image. 
We compare NCAA with a state-of-the-art minimum-volume NMF algorithm that uses a \textit{logdet} penalty to penalize the volume of the columns 
of $X$~\cite{fu2016robust}: 
\begin{equation}
\label{MinVol}
\underset{\substack{W \geq 0 \\ H(:,j) \in \Delta^r  \text{ for all } j}}{\min} \Vert X-WH \Vert_F^2+ \tilde{\lambda} \log \det(W^{T}W + \delta I_r) 
\end{equation}
\noindent where $I_r$ is the identity matrix of size $r$, 
$\tilde{\lambda}$ a regularization parameter and $\delta$ a small scalar constant. This model was shown to be the most efficient compared to other minimum-volume algorithms in \cite{ang2019algorithms}. 
We use the implementation from~\cite{leplatminimum} where it is recommended to use $\tilde{\lambda} = \lambda \frac{||X-WH||_F^2}{\log \det(W^{T}W + \delta I_r)}$ 
with $\lambda = 0.01, 0.1$ to balance the two terms in the objective function, 
and where the initial matrices $(W,H)$ are computed by SNPA. 
The performance metric considered is the average mean removed spectral angle (MRSA) over all couples of corresponding estimated and expected basis vectors, after a proper assignment with the Hungarian algorithm~\cite{kuhn1955hungarian}. 
The MRSA between two vectors $x$ and $y$ is given by $\text{MRSA}(x,y) = \frac{100}{\pi} \text{arcos} \Big(\frac{\langle x - \overline{x}, y - \overline{y}\rangle}{\Vert x - \overline{x}\Vert_{2}\Vert y - \overline{y}\Vert_{2}}\Big) \in \left[  0, 100\right]$  where  $ \langle \cdot, \cdot\rangle $ indicates the scalar product of two vectors and $\overline{\cdot}$ is the mean of a vector.

\subsection{Synthetic data sets}

We first compare the methods on synthetic data sets to investigate the influence of different data distributions. For NCAA, we use SNPA to generate $Y$ with $d=10r$. Moreover, only Algorithm~\ref{algo1} is used, as the purity level will be identical over all the basis vectors and the improvement brought by the fine tuning stage was negligible.  
 
\begin{table} [h]
\centering
\resizebox{\columnwidth}{!}{
\begin{tabular}{c | c | c | c | c }
($\textit{p}$, $\textit{r}$, \textit{$\upsilon$}) & NCAA & MinVolNMF ($\lambda=0.01$) & MinVolNMF ($\lambda=0.1$) & SNPA\\
\hline

($\textit{0.7}$, $7$, $0$) & $\bm{1.13} \pm 2.61$ ($24$) & $7.42 \pm 5.22$ ($0$) & $6.09 \pm 5.30$ ($1$) & $15.04 \pm 2.40$ ($0$)\\
($\textit{0.8}$, $7$, $0$) & $\bm{0.37} \pm 0.61$ ($24$) & $1.99 \pm 2.27$ ($0$) & $1.70 \pm 2.25$ ($1$)& $7.40 \pm 1.20$ ($0$)\\
($\textit{0.9}$, $7$, $0$) & $\bm{0.21} \pm 0.07$ ($20$) & $0.45 \pm 0.23$ ($0$) & $0.41 \pm 0.23$ ($5$)& $3.13 \pm 0.28$ ($0$)\\
($\textit{1}$, $7$, $0$) & $2.12 \cdot 10^{-3} \pm 4.27 \cdot 10^{-3}$ ($8$) & $3.18 \cdot 10^{-3} \pm 6.56 \cdot 10^{-3}$ ($0$) & $3.17 \cdot 10^{-3} \pm 6.53 \cdot 10^{-3}$ ($0$)& $\bm{1.22} \cdot 10^{-5} \pm 1.40 \cdot 10^{-5}$ ($17$)\\
\hline
($0.8$, $\textit{3}$, $0$) & $1.88 \pm 1.05$ ($10$) & $1.73 \pm 0.90$ ($1$) & $\bm{1.47} \pm 0.95$ ($14$) & $7.16 \pm 0.80$ ($0$)\\
($0.8$, $\textit{7}$, $0$) & $\bm{0.37} \pm 0.61$ ($24$) & $1.99 \pm 2.27$ ($0$) & $1.70 \pm 2.25$ ($1$)& $7.40 \pm 1.20$ ($0$)\\
($0.8$, $\textit{12}$, $0$) & $\bm{3.81} \pm 3.97$ ($23$) & $5.70 \pm 3.80$ ($0$) & $5.44 \pm 3.80$ ($2$)& $10.08 \pm 2.53$ ($0$)\\
($0.8$, $\textit {20}$, $0$) & $\bm{6.39} \pm 2.41$ ($22$) & $7.20 \pm 2.48$ ($0$) & $7.09 \pm 2.47$ ($3$)& $10.45 \pm 1.79$ ($0$)\\
\hline
($0.8$, $7$, $\textit{0}$) & $\bm{0.37} \pm 0.61$ ($24$) & $1.99 \pm 2.27$ ($0$) & $1.70 \pm 2.25$ ($1$) & $7.40 \pm 1.20$ ($0$)\\
($0.8$, $7$, $\textit{0.01}$) & $2.31 \pm 3.11$ ($7$) & $2.44 \pm 3.07$ ($0$) & $\bm{2.16} \pm 3.03$ ($18$)& $7.85 \pm 1.98$ ($0$)\\
($0.8$, $7$, $\textit{0.05}$) & $6.32 \pm 2.20$ ($6$) & $6.48 \pm 2.50$ ($0$) & $\bm{5.59} \pm 2.39$ ($19$)& $10.35 \pm 2.94$ ($0$)\\
($0.8$, $7$, $\textit{0.1}$) & $\bm{8.44} \pm 1.73$ ($14$) & $11.02 \pm 3.78$ ($0$) & $9.43 \pm 3.56$ ($11$) & $12.18 \pm 2.16$ ($0$)\\
($0.8$, $7$, $\textit{0.2}$) & $\bm{13.87} \pm 3.30$ ($22$) & $23.23 \pm 3.90$ ($0$) & $21.19 \pm 4.12$ ($1$)& $18.79 \pm 2.29$ ($2$)\\
\end{tabular}
}
\caption{Comparison of the performances of NCAA, MinVolNMF and SNPA on synthetic data, with $n=1000$, $m=10$, $d=10r$ in function of the purity level, rank and noise level respectively in terms of average MRSA over $25$ randomly generated true factors. For each configuration, the best average MRSA is highlighted in bold and the number of times each algorithm performs the best is written in parentheses. }
\label{MyTable}
\end{table} 
 
We generate the data matrices $X \in \mathbb{R}^{m \times n}_+$ as follows. 
We fix $n=1000$ and $m=10$. 
Given the factorization rank $r$, the purity level $p \in (0,1]$  
and the noise level $\upsilon$, we generate 25 random matrices $X = W_tH_t + N$ as follows. 
Each entry of $W_t \in \mathbb{R}^{m \times r}_+$ is drawn from a uniform distribution over the interval $[0,1]$. 
Then, each column is normalized so that its entries sum to one such that all the basis vectors belong to $\Delta^m$.
 The matrix $H_t \in \mathbb{R}^{r \times n}_+$ is generated through a Dirichlet distribution of parameter $\alpha \in \mathbb{R}^{r}$, $\alpha_i=0.05$ for all $i$. Each column is resampled until every entry is smaller than the given purity level $p$. 
 Finally, noise is added to the data such that 
\[
X = \max\left( 0, \tilde{X} + \upsilon ||\tilde{X}||_F \frac{N}{||N||_F} \right),
\]   
where $\tilde{X}=W_t H_t$ and each entry of $N$ follows a Gaussian distribution of mean 0 and standard deviation 1.  

The following values of the variable parameters are considered: 
$r$ $=$ $3$, $7$, $12$, $20$, 
$p$ $=$ $0.8$, $0.9$, $1$ and $\upsilon$ $=$ $0$, $0.01$, $0.05$, $0.1$, $0.2$.

Table~\ref{MyTable} displays the average MRSA and the corresponding standard deviations obtained with the 25 different generated true factors, computed for both NCAA and MinVolNMF (Eq. \eqref{MinVol}) with two values of $\lambda$ ($0.01$ and $0.1$) as well as SNPA. 
Table~\ref{MyTable} also provides the number of times each algorithm returned the best solution (that is, smallest MRSA among the four algorithms). 
We only present the results for some important configurations: we fix $r=7$, $p=0.8$ and $\upsilon = 0$ and, for these values, we vary $r$, $p$ and $\upsilon$ independently.  

We observe the following: 

\begin{itemize} 
\item The variability of the settings generates in general high standard deviations. However, the ranking trend given  by the average MRSA is confirmed by the distribution of the best instances. 
 
\item The MRSA of NCAA is in most cases lower than the one of MinVolNMF. 
Note that MinVolNMF with the two values of $\lambda$ give similar results. 
The baseline SNPA is only competitive in separable cases (that is, when  $p=1$). NCAA performs particularly well in the  difficult scenario when $r > m$, in presence of heavy noise and in highly mixed situations ($p \ll 1$). As opposed to MinVolNMF, NCAA uses the data points to construct the basis vectors hence is more robust in these difficult scenarios. 
\end{itemize}

\subsection{Blind hyperspectral unmixing}
\label{sec:blindhu}

\begin{center}
\begin{figure}
\begin{center}
\includegraphics[scale=5]{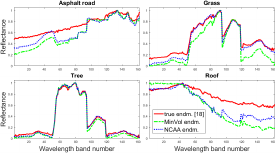}
\end{center}
\captionof{figure}{Endmembers comparison between NCAA, MinVolNMF and the ground truth for Urban image with $r=4$.}
\label{endmembers}
\end{figure}
\end{center}

\begin{figure}
\centering 
\begin{subfigure}{.45\textwidth}
  \centering
  \includegraphics[width=\linewidth]{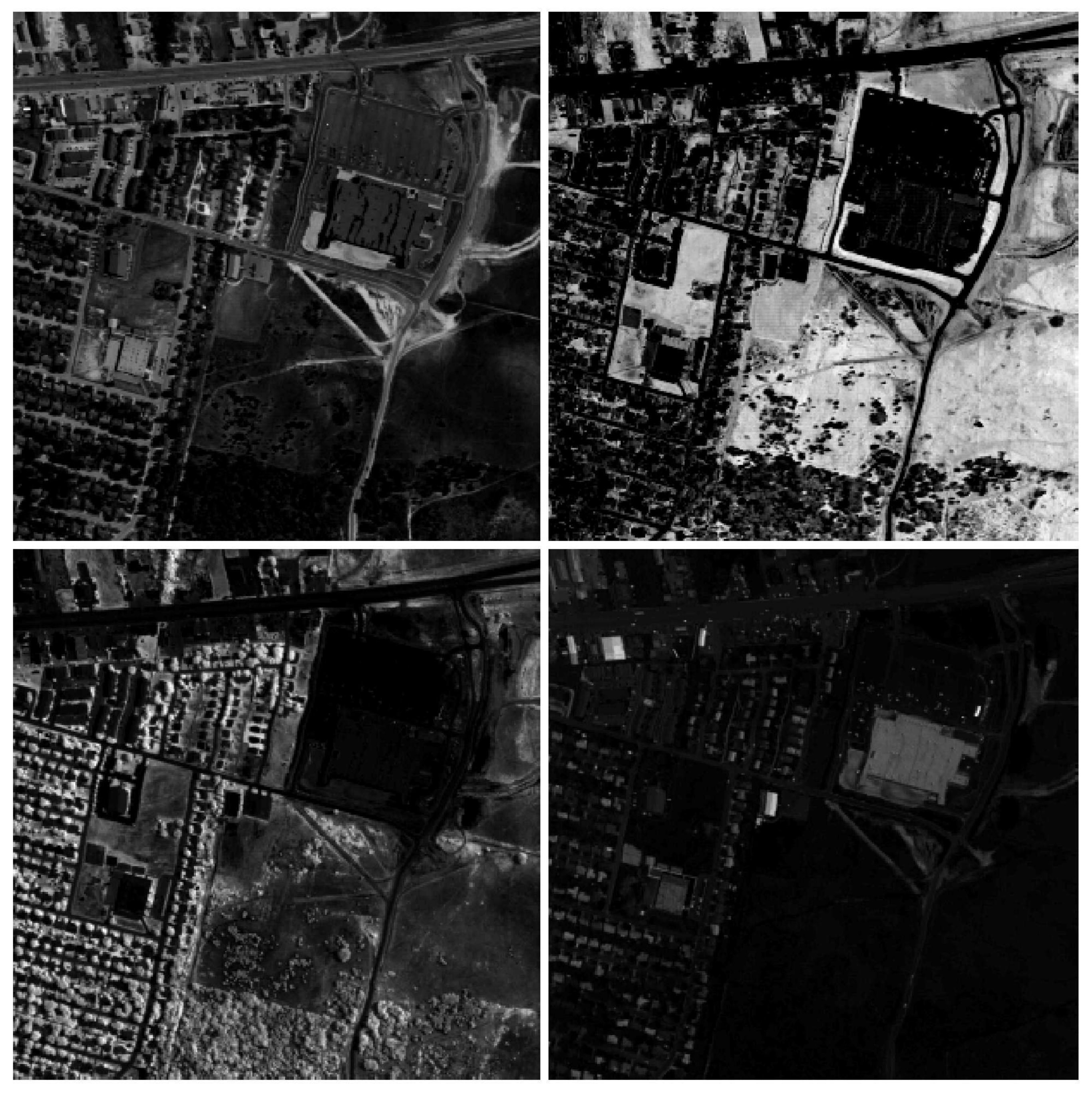}
  \caption{MinVolNMF}
  \label{MVA}
\end{subfigure}
\quad 
\begin{subfigure}{.45\textwidth}
  \centering
  \includegraphics[width=\linewidth]{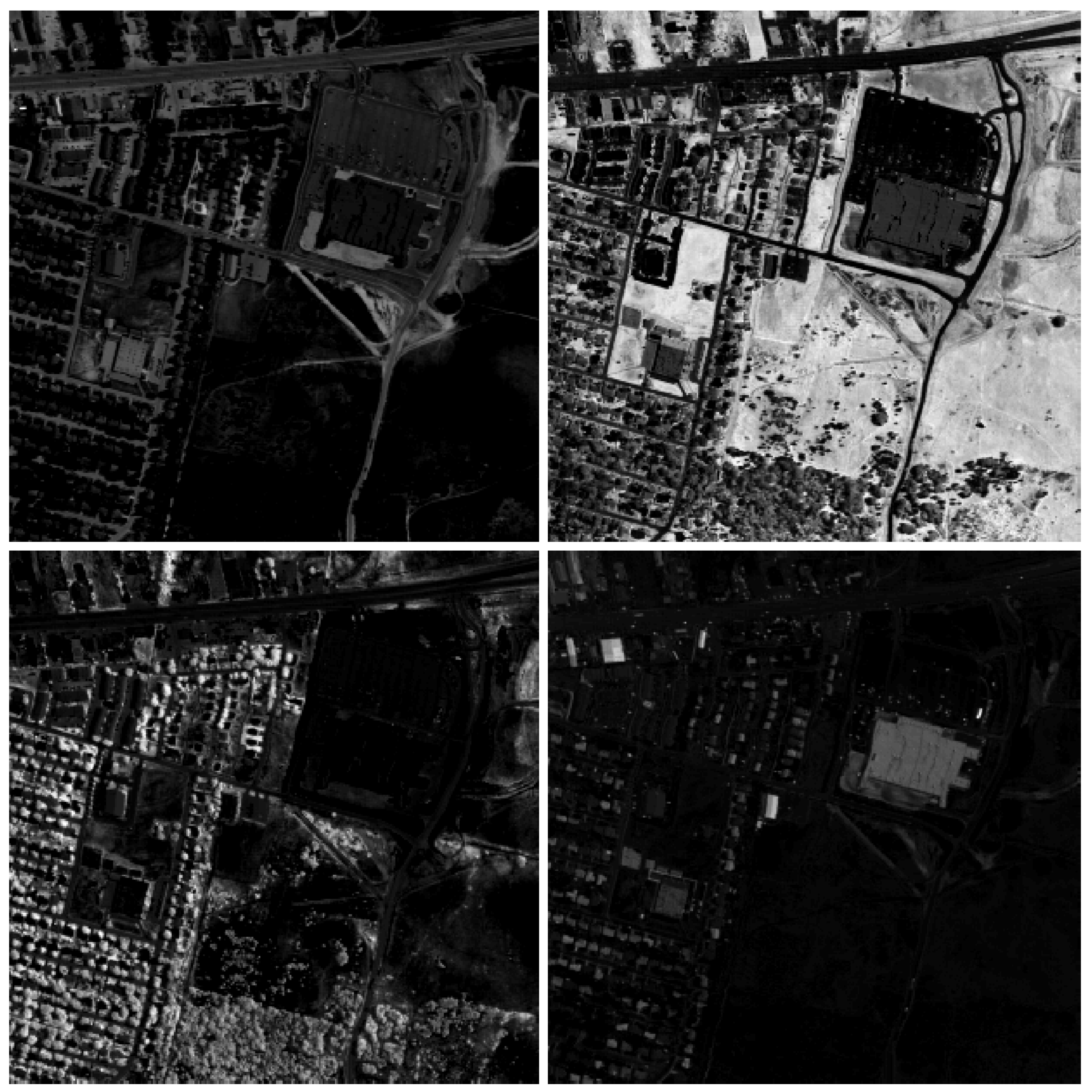}
  \caption{NCAA}
  \label{NCAA}
\end{subfigure}
\caption{Material abundances for Urban image with $r=4$. From left to right, on top: road, grass; on bottom: tree, roof.}
\label{fig:test}
\end{figure}

Hyperspectral unmixing consists in identifying $r$ materials, or endmembers, inside a hyperspectral image made of $n$ pixels in $m$ spectral bands. The HYDICE Urban image  is made of $n=307 \times 307$ pixels in $m=162$ denoised spectral bands. Four important materials are asphalt road, grass, tree and roof, and we fix $r=4$ (see for example \cite{zhu2017spectral} for more details). 
The matrix $Y$ is set up with HC, and we use $d=20$.
On Fig.~\ref{endmembers}, a comparison of the normalized spectral signatures of the endmembers obtained with both NCAA (including the fine-tuning step) and MinVolNMF, and the ground truth is presented. We observe that NCAA produces spectral signatures very close to the ground truth. Besides, the MRSA of our model is $5.56$  while the one of MinVolNMF is $5.73$. Moreover, the abundance maps on Fig.~\ref{NCAA} show that NCAA is able to retrieve meaningful proportions of each endmember in the initial image. For comparison, the abundance maps of MinVolNMF are presented on Fig.~\ref{MVA}. 

\section{Conclusion} \label{conclusion}


In this letter, we have proposed a new NMF model called near-convex archetypal analysis (NCAA) which on the one hand guarantees a low approximation error and on the other hand is interpretable like archetypal analysis from which it is inspired. The value of the parameter $\epsilon$ in NCAA~\eqref{my_model} plays the role of a cursor regulating the maximum distance between the basis vectors and the convex hull of the data points $X$.  
Although it is possible to estimate the value of the parameter $\epsilon$ that retrieves the true basis vectors in simple settings, we would be interested in analysing the uniqueness of the solution of NCAA. 
As the intuition of NCAA is close to the one of minimum-volume NMF, it would be particularly interesting to extend the identifiability 
results obtained in~\cite{huang2013non, lin2015identifiability, fu2018identifiability}.  
It would also be interesting to explore other models inspired by NCAA. For example, using $Y=X$ and imposing row sparsity of $A$ would make the model learn $Y$ and the number of points $d$ automatically; see~\cite{bauckhage2009making} for a similar idea.  
Other models using regularization terms in the objective function rather than fixing hard constraints on $A$ could also be interesting to explore. 




%
%
%
%

\bibliographystyle{abbrv}
\bibliography{our_biblio}

\end{document}